\documentclass[preprint2]{aastex}

\usepackage{epsfig}
\usepackage{psfig}
\usepackage{lscape}

\newcommand{\chisq}{$\chi^{2}$}
\newcommand{\G}{G308.3-1.4}

\slugcomment{}

\shorttitle{Chandra observation of \G}
\shortauthors{Hui et al.}

\begin{document}

\title{Identification campaign of supernova remnant candidates 
in the Milky Way $-$ I:  Chandra observation of \G}

\author{C. Y. Hui\altaffilmark{1}, K. A. Seo\altaffilmark{1},
R. H. H. Huang\altaffilmark{2}, L. Trepl\altaffilmark{3}, 
Y. J. Woo\altaffilmark{1}, 
T.-N. Lu\altaffilmark{2}, A. K. H. Kong\altaffilmark{2,4},
and F. M. Walter\altaffilmark{5}}

\altaffiltext{1}{Department of Astronomy and Space Science, 
Chungnam National University, Daejeon 305-764, Korea}
\altaffiltext{2}
{Institute of Astronomy and Department of Physics, National 
Tsing Hua University, Hsinchu, Taiwan}
\altaffiltext{3}
{Astrophysikalisches Institut und Universit\"{a}ts-Sternwarte, 
Universit\"{a}t Jena, Schillerg\"{a}$\beta$chen 2-3, 07745 Jena, 
Germany}
\altaffiltext{4}
{Golden Jade Fellow of Kenda Foundation, Taiwan}
\altaffiltext{5}
{Department of Physics and Astronomy, Stony Brook University, Stony Brook, NY 11794-3800, USA}


\begin{abstract}
ROSAT all-sky survey (RASS) data have provided
another window to search for supernova remnants
(SNRs). In reexamining this data archive, a list of unidentified 
extended X-ray objects have been suggested as promising 
SNR candidate. However, most of these targets have not yet been
fully explored by the state-of-art X-ray observatories. 
For selecting a pilot target for a long-term identification 
campaign, we have observed the brightest candidate, \G, with  
Chandra X-ray observatory. An incomplete shell-like X-ray structure which 
well-correlated with the radio shell emission at 843~MHz has been revealed. 
The X-ray spectrum suggests the presence of a shock-heated plasma. All these
evidences confirm \G\ as a SNR. 
The brightest X-ray point source detected in this field-of-view is also the one locates 
closest to the geometrical center of \G, which has a soft spectrum. The intriguing 
temporal variability and the identification of optical/infrared counterpart rule out 
the possibility of an isolated neutron star. On the other hand, the spectral energy 
distribution from $K_{s}$ band to $R$ band suggests a late-type star. Together 
with a putative periodicity of $\sim1.4$~hrs, the interesting 
excesses in $V$, $B$ bands and H$\alpha$ suggest this source as a promising 
candidate of a compact binary survived in a supernova explosion (SN). 
\end{abstract}


\section{INTRODUCTION}
Most Galactic SNRs were identified by radio observations 
(Green 2009). However, one should notice
that there are various selection effects in the Galactic surveys for 
SNRs in the radio band. For example, Schaudel (2003)
suggested that Galactic SNRs older 
than $\sim5\times10^{4}$~yrs are difficult to be detected 
with the current radio telescopes due to their low surface brightness.  
Therefore, the currently known SNR population of our Galaxy is certainly 
biased.

There are 274 SNRs have been uncovered in the Milky Way so far (Green 2009). 
This known sample is far smaller than the expected population. Assuming 
a typical evolution timescale of SNRs before they merge with the ISM 
($\sim10^{5}$~yrs) and a event rate of 2~SNe/century in the Milky Way 
(Dragicevich et al. 1999), $\sim2000$~SNRs are expected in our Galaxy.
Such a great deficit again points to the selection effects in 
the past surveys. 

Besides radio observations, the ROSAT all-sky survey (RASS) provides 
us with an alternative window for searching
SNRs in our Galaxy. Schaudel (2003) suggests that the 
energy band (0.1$-$2.4~keV) of the Positional Sensitive Proportional Counter (PSPC) 
on board ROSAT allows 
the detection of SNRs for an interstellar absorption
$\leq3\times10^{22}$~cm$^{-2}$. 
A simulation of the theoretical distribution of SNe and
their remnants in the Galactic plane has predicted somewhat 
more than 200 SNRs should possibly be detected in RASS
(Busser 1998). In examining the RASS database, Schaudel (2003) 
has reported  $\sim100$ unidentified extended X-ray sources as 
the possible SNR candidates. 

Although the RASS database allows one to search 
for SNR candidates, the short exposure of few
hundred seconds and the poor spatial resolution 
in survey mode ($\sim96''$) prevent any firm
identification of their nature. In addition, 
a possible hard X-ray ($>2$ keV) component could arise from
the interactions of the reflected shocks with the 
dense ambient medium or alternatively from the
synchrotron emission radiated by the relativistic leptons. 
However, the limited 
energy bandwidth of ROSAT does not allow
one to determine whether an additional component 
presents in the hard X-ray band.
In view of its much improved spatial resolution and 
the enlarged effective area, Chandra X-ray Observatory 
and XMM-Newton Observatory  
provide the most suitable instruments to clarify 
the emission nature of these candidates. And this  
motivates an extensive identification campaign of all 
these unidentified extended RASS objects with the 
state-of-art X-ray telescopes. 

\begin{figure*}[b]
\centerline{\psfig{figure=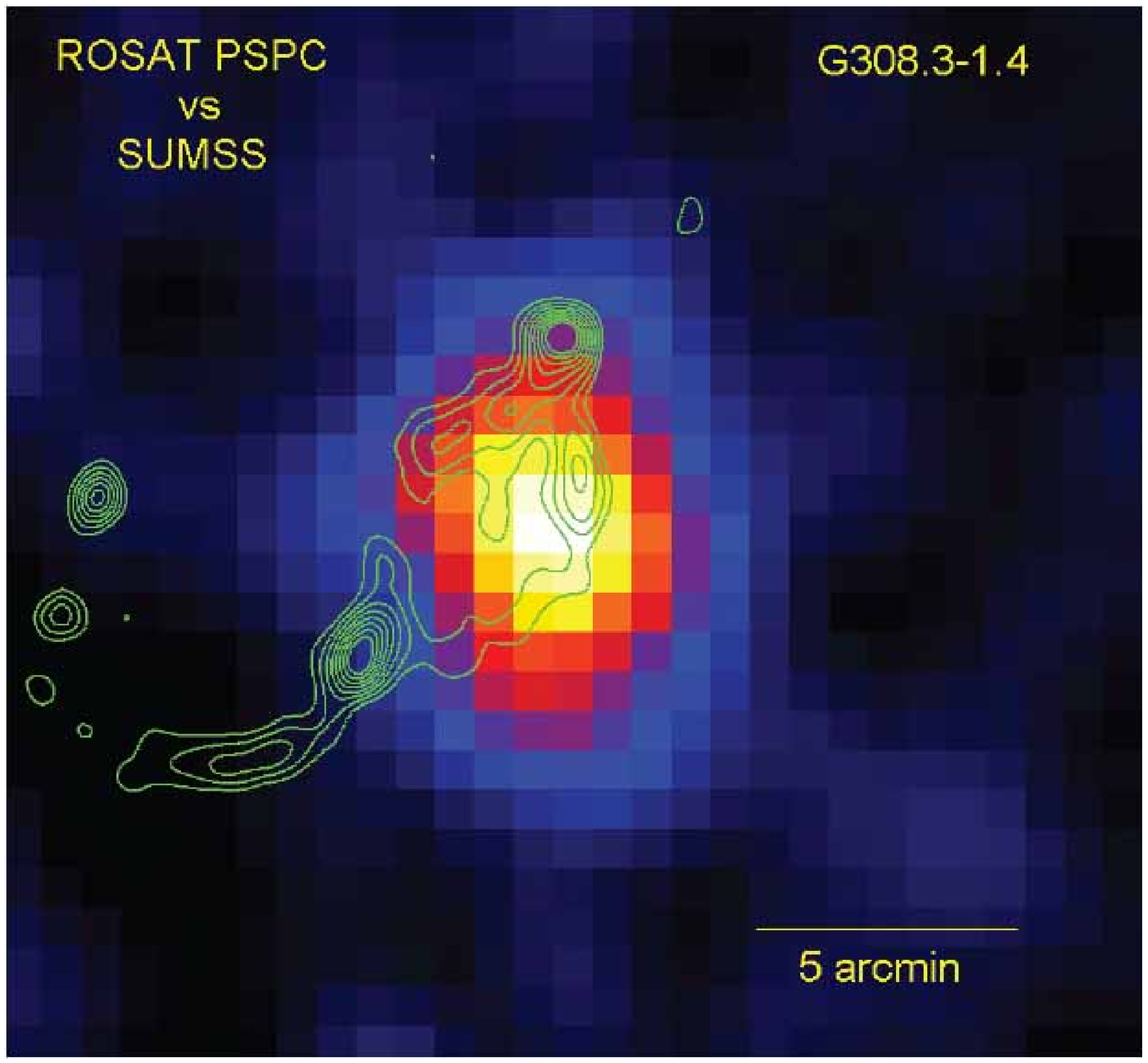,width=16cm,clip=,angle=0}}
\caption[]{RASS image of \G\ in the energy band of $0.1-2.4$~keV 
with radio contour lines at the levels of $5-30$~mJy/beam from the
SUMSS data overlaid. Top is north and left is east.}
\label{rass}
\end{figure*}

For an initial stage of the campaign, we chose the brightest 
SNR candidate in the list provided by Schaudel (2003), \G, as
the pilot target. \G\ was firstly listed as a possible SNR candidate 
in the MOST SNR catalogue based on its radio morphology and the 
spectral index (Whiteoak 1992). 
In the RASS image, \G\ shows centrally-peaked X-ray emission 
with a diameter of $\sim10'$ which coincides with a radio 
arc-like feature (see Figure~\ref{rass}). 
The radio contours obtained from the 843~MHz Sydney University Molonglo Sky 
Survey {\bf (SUMSS)} data (Bock et al. 1999). 
However, owing to the poor spatial resolution of RASS data, it 
is unclear whether the X-rays are originated from the radio shell 
structure.
Also, the limited photon statistic of \G\ in the 
RASS data do not allow any further probe of its emission properties. 
In this paper, we report the first detailed investigation for the 
nature of \G\ with Chandra observation.  

\section{OBSERVATIONS}

\G\ was observed with Chandra in 2010 June $26-27$ with the Advanced CCD Imaging
Spectrometer (ACIS) using a frame time of 3.2~s. Since the source extent cannot be
accurately determined from the RASS data, we utilized the whole ACIS-I
CCD array with an intention to cover the whole feature with aimpoint at the nominal 
center of the extended RASS feature 
(i.e. RA=13$^{\rm h}$40$^{\rm m}$56.3$^{\rm s}$ Dec=-63$^{\circ}$43$^{'}$32.6$^{"}$ 
(J2000)). For data reduction as well as analysis, we utilized the 
{\bf C}handra {\bf I}nteractive {\bf A}nalysis of {\bf O}bservations software 
(CIAO~4.3) throughout this study. We have firstly reprocessed the data with CALDB 
(ver.~4.4.3) to correct for an error in the time-dependent gain correction (TGAIN) 
during the observation by using the script \emph{chandra\_repro}. To facilitate 
a source detection with high positional accuracy, we have applied the 
subpixel event repositioning in reprocessing the data. The effective exposure of 
this observation is $\sim14.9$~ks. We restricted all the analysis in an energy band 
of $0.5-8$~keV. 

We have also conducted a series of
Target-of-Opportunity (ToO) observations of this source with
the {\bf U}ltra-{\bf V}iolet and {\bf O}ptical {\bf T}elescope (UVOT) onboard the SWIFT satellite.
The pre-processed calibrated science grade (level II) data were used. This pre-processing includes bad pixel
correction and the reduction of the modulo 8 fixed-pattern which results from the fact that raw
image pixel are smaller by up to factor of 8 than the original detector pixels. In addition, the
images have been flatfield corrected. The data were then analyzed by using the HEASoft Swift package.

\section{SPATIAL ANALYSIS}

\begin{figure*}[b]
\centerline{\psfig{figure=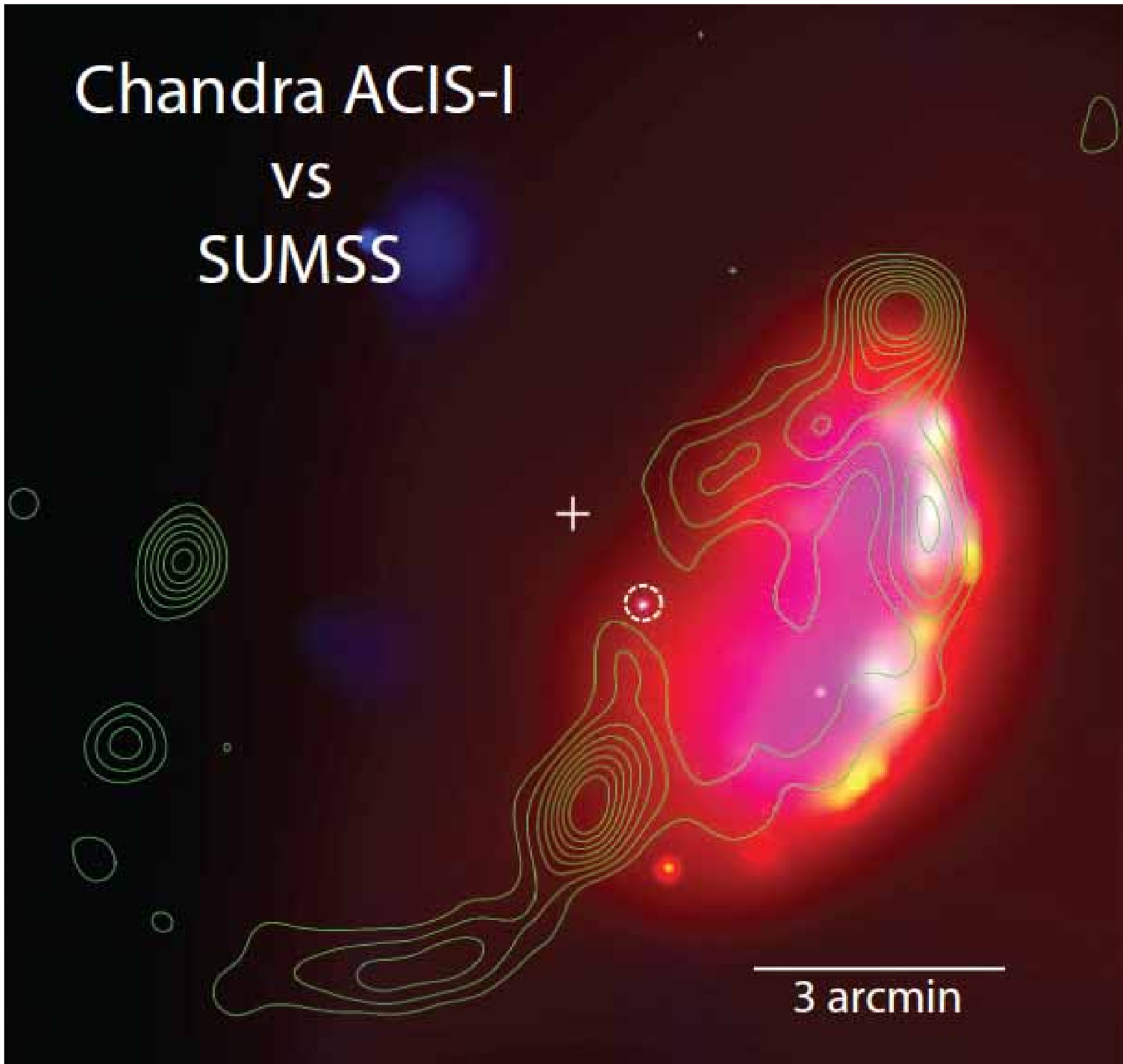,width=16cm,clip=,angle=0}}
\caption[]{$10^{'}\times10^{'}$ Chandra ACIS-I X-ray colour image of 
\G\ (red: $0.5-1$~keV, green: $1-2$~keV, blue: $2-8$~keV). The 
binning factor of this image is $2^{"}$. Adaptive smoothing has been 
applied to achieve a minimum signal-to-noise ratio of 3. The same 
radio contour lines as in Fig.~\ref{rass} are overlaid for comparison. 
The geometrical center inferred from the X-ray morphology is illustrated
by the cross. A bright source which locates closest to the geometrical center 
(i.e. source\#7 in Fig.~\ref{wavelet} and Tab.~\ref{x_src}) is highlighted by the dash circle.
Top is north and left is east.}
\label{rgb}
\end{figure*}

An X-ray color image of the ACIS-I data of \G\ is displayed in Figure~\ref{rgb}. 
With the superior spatial resolution of Chandra, an incomplete X-ray shell 
structure have been revealed. We have overlaid the same SUMSS radio contours 
(cf. Fig.~\ref{rass}) onto Figure~\ref{rgb}. This comparison demonstrates a 
clear correlation between the X-ray emission and the radio feature, and hence 
their connection is confirmed unambiguously and suggest \G\ belongs to 
category of mixed-morphology SNRs. The incomplete X-ray shell-like morphology 
conforms with a circle with an angular radius of $\theta\sim5^{'}$ 
approximately centered 
at RA=13$^{\rm h}$41$^{\rm m}$32$^{\rm s}$ Dec=-63$^{\circ}$42$^{'}$44$^{"}$
(J2000). 

Figure~\ref{rgb} also shows the hardness distribution of the X-rays from \G. 
Several components can be identified in this color image. 
Apart from the bright outer rim emission, a fainter but harder feature is noticed 
in its inner edge which we refer as the ``inner rim" in this paper. Besides these 
two shell-like components, soft and fainter diffuse emission is found in the region
close to the center. 

\begin{figure*}[b]
\centerline{\psfig{figure=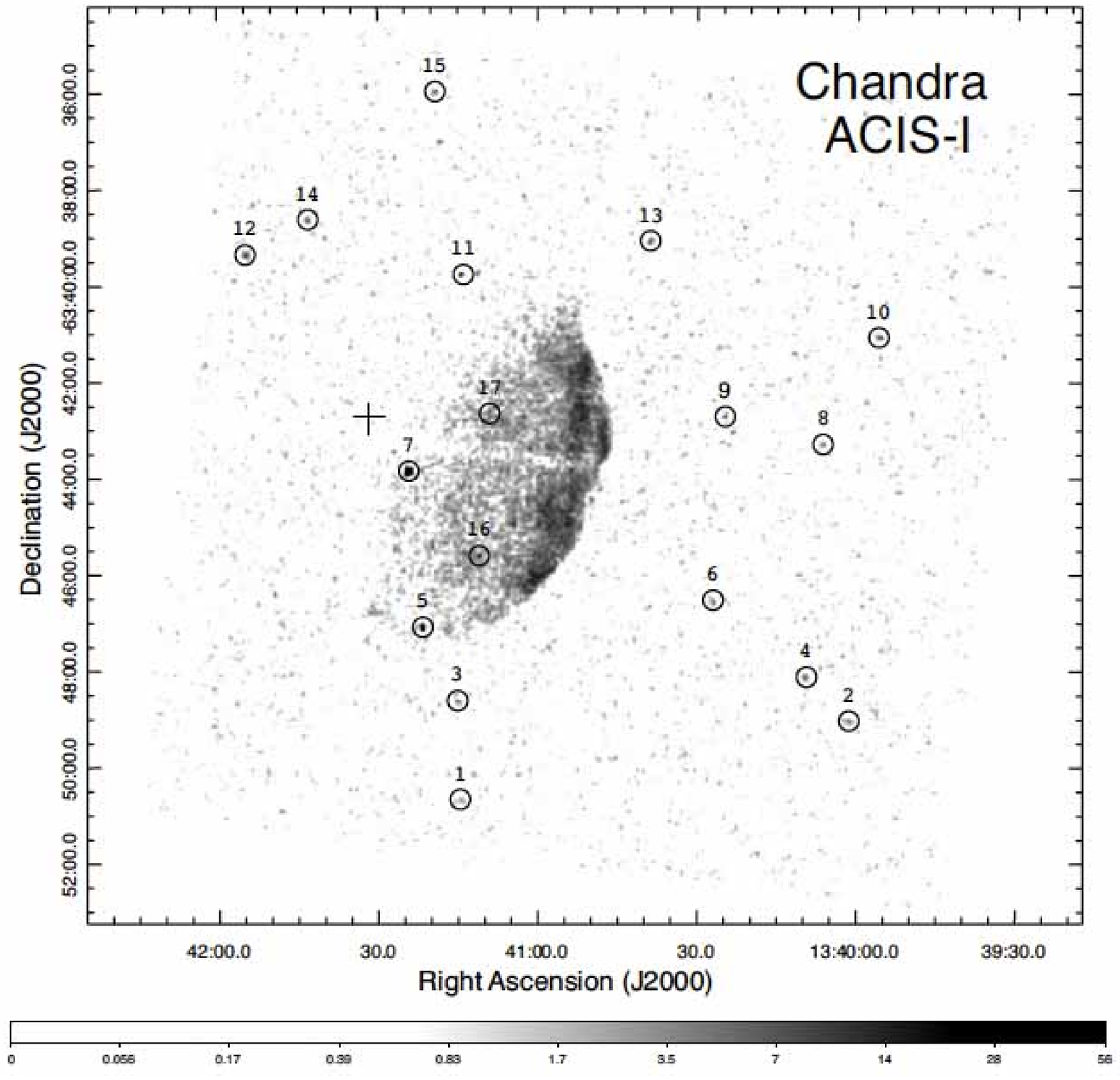,width=16cm,clip=,angle=0}}
\caption[]{17 X-ray sources detected in the whole ACIS-I image of \G\ by wavelet 
detection algorithm. The properties of these source are summarized in Tab.~\ref{x_src}.
The geometrical center inferred from the X-ray morphology is illustrated
by the cross. 
}
\label{wavelet}
\end{figure*}

This observation also enable a search for X-ray point sources in the vicinity of \G.  
By means of a   
wavelet source detection algorithm, 17 sources has been detected in the field-of-view. 
These sources are labeled in Figure~\ref{wavelet}. The source positions, 
positional errors, net count rates, signal-to-noise ratios as well as the estimates of 
spatial extent of all these sources are given in Table~\ref{x_src}. 
The brightest object is the source marked as \#7 in Figure~\ref{wavelet}, which 
is also the one located closest the geometrical center. Cross-correlating these
sources with the SIMBAD and NED databases did not result in any identification
within a search radius of 5 arcsec around each source.

To investigate if these sources are promising isolated neutron star candidates we proceeded to
search for their possible optical counterparts by utilizing the USNO-B1.0 catalogue 
(Monet et al. 2003). The X-ray-to-optical flux ratio, 
$f_{x}/f_{\rm opt}$, provides a rudimentary parameter 
for discriminating the source nature. For an isolated neutron star,  
$f_{x}/f_{\rm opt}$ is typically larger than 1000 (cf. Haberl 2007). On the other hand,
$f_{x}/f_{\rm opt}$ for the field stars and the active galactic nuclei are much lower
which typically $<0.3$ and $<50$ respectively (Maccacaro et al. 1988; Stocke et al. 1991).
We have systematically searched for the optical counterparts within a
$10\sigma\times10\sigma$ error box centered at each X-ray position. Among all 17 sources, 
only the object labeled as \#5 have no optical counterpart found in its error box. 
For those 16 X-ray objects have optical counterparts identified, we have calculated the
X-ray-to-optical flux ratios with the $B-$band magnitudes from the USNO-B1.0 catalogue. The
X-ray fluxes $f_{x}$ of these objects are estimated from the net count rates in Table~\ref{x_src} with the 
aid of {\it PIMMS} (ver. 4.3) by assuming a power-law spectrum with a photon index of 2. The flux ratio 
of source \#7, $f_{x}/f_{B}\sim1.5$ is found to be the highest among all these 16 sources. 
For source \#5, despite no optical counterpart have been identified in this search, the limiting 
magnitude of USNO-B1.0 catalogue (i.e. $\sim21$) results in a lower limit 
of $f_{x}/f_{B}\gtrsim3$ which is not tight enough to constrain its source nature.  
Furthermore, its X-ray spectral properties do not conform with that of a neutron star (see \S5). 
Therefore, we conclude that we do not find any concrete evidence for an isolated neutron star in 
this investigation. 

\begin{figure*}[t]
\centerline{\psfig{figure=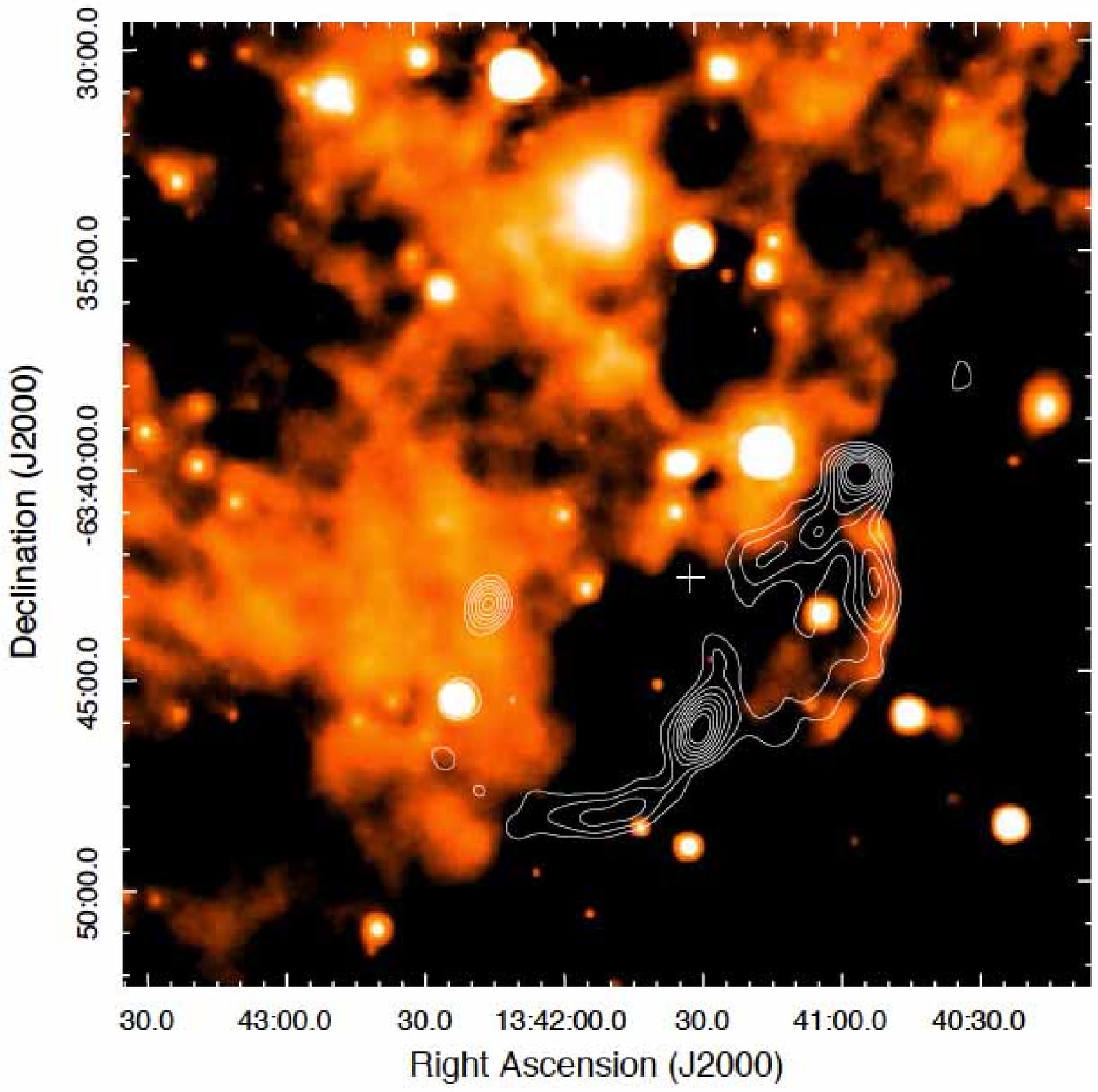,width=18cm,clip=,angle=0}}
\caption[]{The 22~$\mu$m image of the $23'\times23'$ field around \G\ as observed by WISE. 
The same radio contour lines as in Fig.~\ref{rass} and Fig.\ref{rgb} are overlaid for comparison.
The geometrical center inferred from the X-ray morphology is illustrated
by the cross.
}
\label{wise_sumss}
\end{figure*}

Besides radio and X-ray data, we have also explored the infrared data obtained by
{\bf W}ide-field {\bf I}nfrared {\bf S}urvey {\bf E}xplorer (WISE; Wright et al. 2010).
In Figure~\ref{wise_sumss}, we compared the 22~$\mu$m image with the same set of radio contours
used in Figure~\ref{rass} and Figure~\ref{rgb}. An extended incomplete elliptical feature has been
identified in this mid-infrared image. This feature has a semi-major/semi-minor axis of $\sim10'/8'$ with the
major axis oriented $\sim20^{\circ}$ from north. On the southwestern edge of this feature, an infrared rim emission
is found to be well-correlated with the radio shell and the outer-rim X-ray emission of \G\
(cf. Figure~\ref{rgb}). This strongly suggests that these rim structures seen in different wavelengths
are intrinsically related.

Apart from this rim, there appears to be a cavity in the southwestern quarter of this infrared
feature (Fig.~\ref{wise_sumss}).
This is the region where the X-ray emission is most prominent (see Fig.~\ref{rgb} and Fig.~\ref{wavelet}).
We also noticed that the incomplete radio shell apparently
complements the peripheral of the incomplete elliptical infrared structure. 
Also, there is no X-ray/radio emission toward the northeastern region where the infrared feature is bright.
Such morphology resembles those of the SNRs which have interactions with the surrounding cloud complex 
(e.g. Sasaki et al. 2004). However, as there is no reliable distance estimates for the X-ray/radio and the infrared 
features, the physical association between the large infrared structure and the X-ray/radio shell remains uncertain.
Further investigations for this possible shock-cloud interaction,
such as maser observations (see Wardle \& Yusef-Zadeh 2002), are encouraged. 

\section{TEMPORAL ANALYSIS}

We have also examined the temporal behavior of these two relatively bright sources 
with reasonable photon statistics. With the aid of the tool \emph{axbary}, their arrival times were 
firstly barycentric-corrected by using the corresponding X-ray positions reported in Table~\ref{x_src}.
With the aid of the tool \emph{glvary}, we searched for variability by using the Gregory-Loredo algorithm 
(Gregory \& Loredo 1992).
For source \#5, a zero variability index has been assigned and the probability of variable signal is found 
to be $\sim11\%$.\footnote{http://cxc.harvard.edu/ciao/threads/variable/index.html\#output} 
And hence, no evidence of variability is found for this source. On the other hand, a 
variability index of 8 is assigned for source \#7 and the probability of variable signal is at the level 
$>99.99\%$, which strongly suggests the presence of temporal variability. 
We have cross-checked these results by adopting other statistical tests 
(e.g. $\chi^{2}$ test), which resulted in the same conclusion as aforementioned. 

\begin{figure*}[t]
\centerline{\psfig{figure=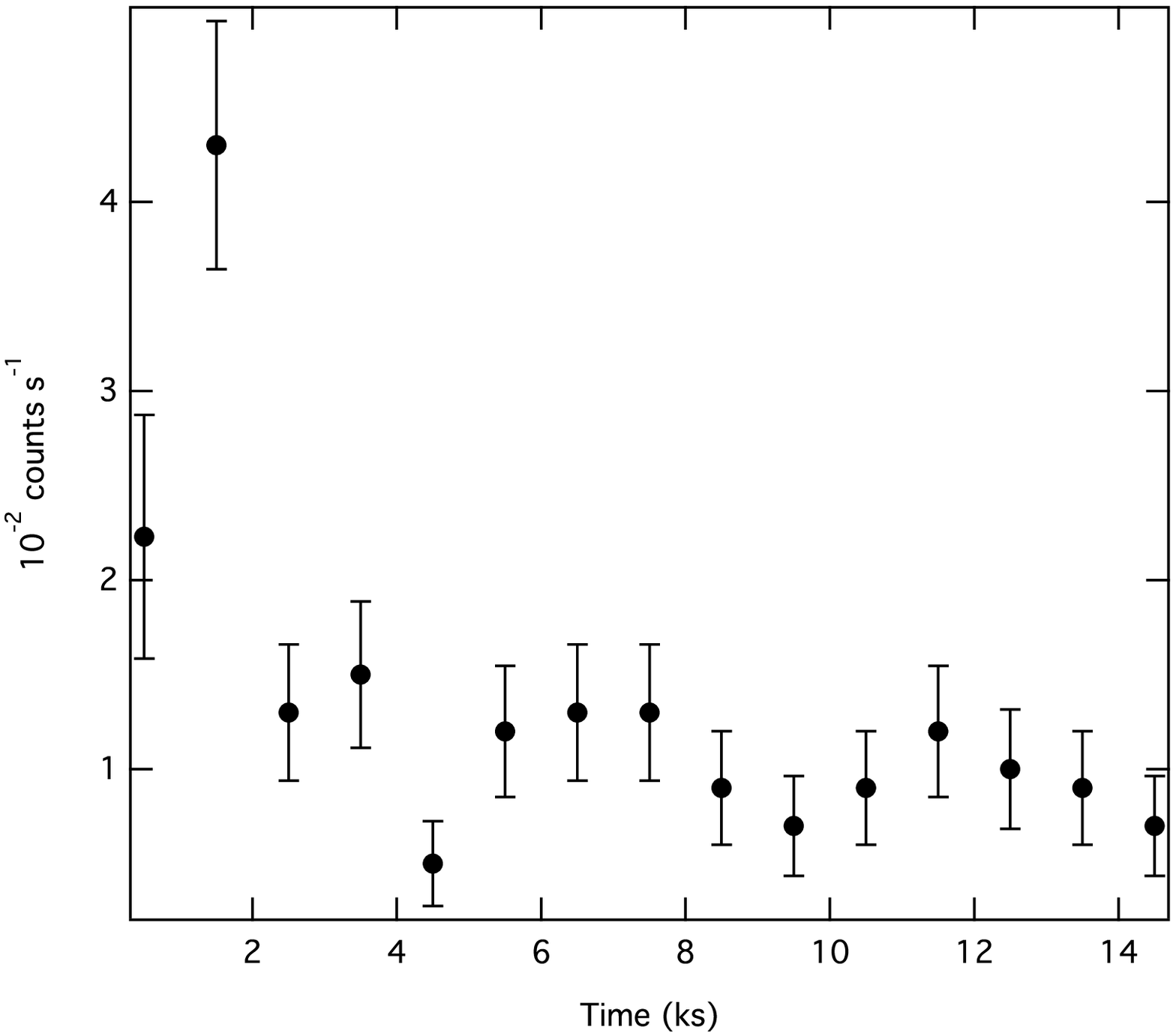,width=16cm,clip=,angle=0}}
\caption[]{The light curve of the X-ray source \#7 detected by Chandra ACIS-I which has shown its 
variability during the observation. The error bars represent $1\sigma$ uncertainties.}
\label{src7_var}
\end{figure*}

\begin{figure*}[t]
\centerline{\psfig{figure=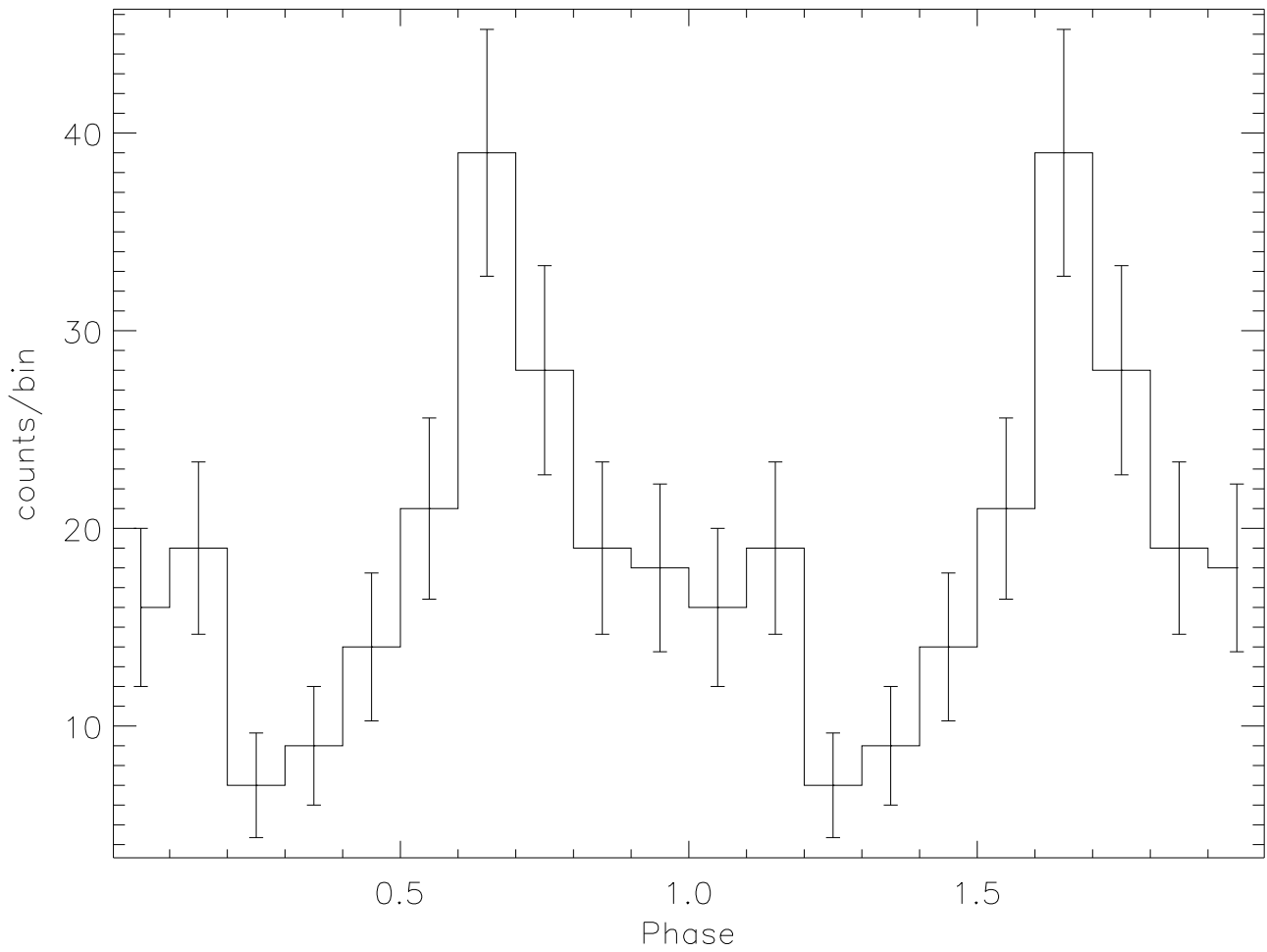,width=14cm,clip=,angle=90}}
\caption[]{X-ray counts of source \#7 versus phase for a periodicity candidate of 
1.4~hrs. Two periodic cycles are shown for clarity. The error bars represent $1\sigma$ uncertainties.}
\label{orbit_fold}
\end{figure*}

Figure~\ref{src7_var} shows the light curve of source \#7 with a binned time of 1000 s. Burst-like activity
is noticed in the beginning of this observation (i.e. $2^{\rm nd}$ bin of Fig.\ref{src7_var}). However, 
the limited photon statistics does not allow a firm conclusion to be drawn. Except for this, the light 
curve appears to be rather steady.
After excluding the time segment possibly contains the flare, 
we proceeded to search for the possible periodic signals from source \#7. 
By fitting the light curve with a sinusoidal, we have identified an interesting periodicity candidate at  
$P=1.4\pm0.2$~hrs. The epoch-folded light curve is shown in Figure~\ref{orbit_fold}. $\chi^{2}$ test 
indicates that it is different from a uniform distribution at $99.97\%$ confidence level. 
While the periodic 
variation appears to be promising, we stress that the peak-to-peak modulation is only at a marginal 
level of $\sim4\sigma$ due to the large error bars. 
A deeper observation is required to confirm this interesting periodicity candidate. 
On the other hand, we do not find any promising periodic signal from source \#5 in 
this observation. 

\begin{table*}[b]
\caption{Positions of the X-ray sources detected in the field of ACIS-I as labeled
in Fig.~\ref{wavelet}. The corresponding $1\sigma$ positional errors, net count rates, 
signal-to-noise ratios as well as the estimates of source extent are tabulated.}
\label{x_src}
\begin{center}
\begin{tabular}{lccccccc}
\hline
\hline
Source & RA (J2000) & Dec (J2000) & $\delta$RA  & $\delta$Dec & Net count rate & S/N$^{a}$ & PSF RATIO$^{b}$ \\ \hline
       & h:m:s & d:m:s & arcsec & arcsec & $10^{-3}$~cts~s$^{-1}$ & $\sigma_{\rm G}$ & \\ \hline
1 & 13:41:15.00 & -63:50:41.65 & 2.97 & 0.82 & 0.85$\pm$0.31 & 3.13 & 2.26 \\
2 & 13:40:01.41 & -63:49:04.98 & 2.81 & 1.25 & 1.45$\pm$0.40 & 4.38 & 2.64 \\
3 & 13:41:15.06 & -63:48:37.41 & 2.74 & 0.82 & 0.92$\pm$0.34 & 3.06 & 4.19 \\
4 & 13:40:10.00 & -63:48:10.30 & 2.05 & 1.45 & 1.29$\pm$0.40 & 3.76 & 3.74 \\
5 & 13:41:22.00 & -63:47:05.85 & 0.25 & 0.19 & 6.47$\pm$0.78 & 13.31 & 1.64 \\
6 & 13:40:27.00 & -63:46:31.78 & 2.78 & 1.03 & 1.15$\pm$0.36 & 3.77 & 7.13 \\
7 & 13:41:24.00 & -63:43:52.40 & 0.37 & 0.19 & 12.96$\pm$1.03 & 26.20 & 5.25 \\
8 & 13:40:06.00 & -63:43:20.13 & 2.15 & 1.32 & 0.71$\pm$0.28 & 2.87 & 3.19 \\
9 & 13:40:24.57 & -63:42:40.36 & 2.03 & 1.45 & 1.32$\pm$0.38 & 4.25 & 10.30 \\
10 & 13:39:56.00 & -63:41:04.22 & 1.77 & 0.77 & 1.46$\pm$0.42 & 4.24 & 1.96 \\
11 & 13:41:14.29 & -63:39:46.21 & 2.01 & 0.46 & 1.61$\pm$0.38 & 6.00 & 4.53 \\
12 & 13:41:55.00 & -63:39:21.22 & 1.18 & 0.68 & 2.14$\pm$0.45 & 6.74 & 1.46 \\
13 & 13:40:38.82 & -63:39:04.23 & 1.13 & 0.68 & 1.95$\pm$0.47 & 5.32 & 3.04 \\
14 & 13:41:42.44 & -63:38:36.12 & 3.08 & 1.01 & 1.19$\pm$0.37 & 3.71 & 3.46 \\
15 & 13:41:19.00 & -63:35:55.49 & 2.87 & 1.11 & 1.06$\pm$0.36 & 3.40 & 2.41 \\
16 & 13:41:11.00 & -63:45:39.36 & 1.23 & 0.56 & 2.45$\pm$0.73 & 3.63 & 12.21 \\
17 & 13:41:08.98 & -63:42:39.32 & 0.96 & 0.55 & 3.19$\pm$0.89 & 3.84 & 11.03 \\
\hline\hline
\end{tabular}
\end{center}
$^{a}$ Estimates of source significance in units of Gehrels error:
$\sigma_{G}=1+\sqrt{C_{B}+0.75}$ where
$C_{B}$ is the background counts.\\
$^{b}$ The ratios between the source extents and the estimates of the PSF sizes.
\end{table*}

\begin{figure*}[b]
\centerline{\psfig{figure=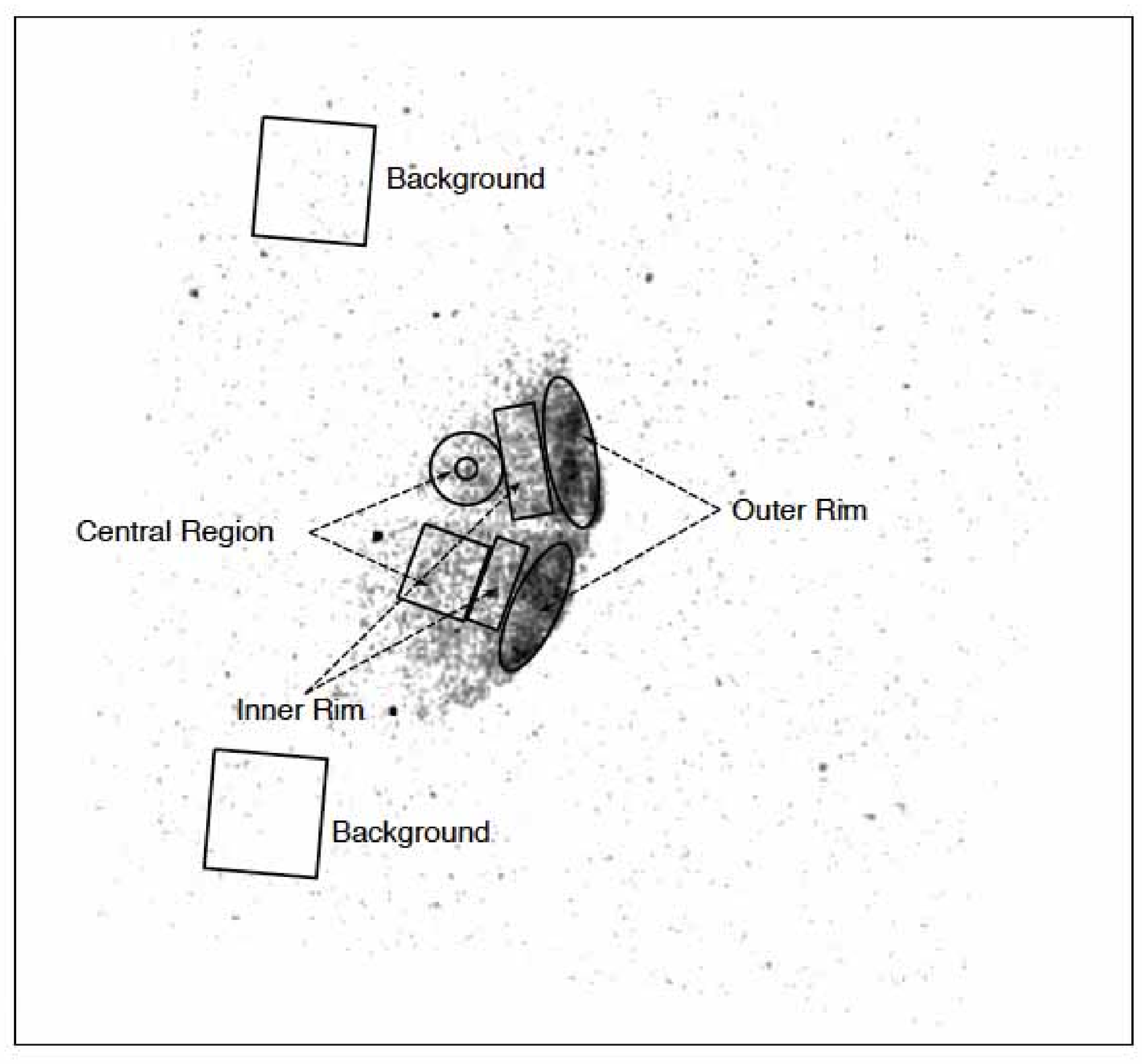,width=16cm,clip=,angle=0}}
\caption[]{Illustration of the regions used to extract the spectra from different 
parts of \G.}
\label{spec_region}
\end{figure*}

\section{SPECTRAL ANALYSIS}

For the spectral analysis, we extracted the spectra for various components of \G\
from the regions illustrated in Figure~\ref{spec_region}. These regions are 
carefully selected so that they are free from the contamination of the point spread
function (PSF) wing from any resolved source. The background 
spectra for the corresponding CCD chips were extracted from the 
source free regions. 

\begin{figure*}
\centerline{\psfig{figure=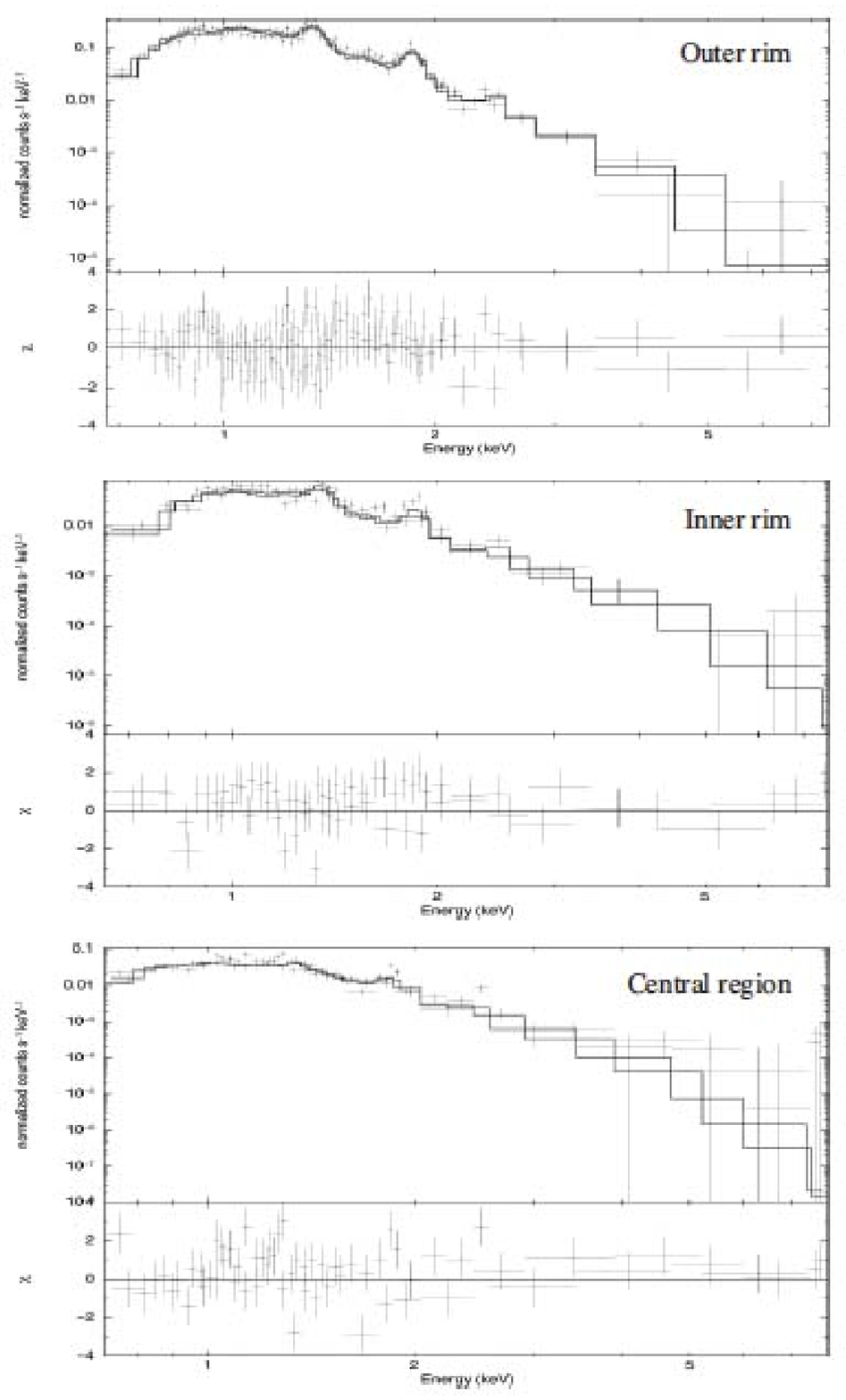,width=12cm,clip=,angle=0}}
\caption[]{X-ray spectra of the emission from various regions of \G\ 
as observed by Chandra ACIS-I. 
Residuals resulted from fitting an absorbed non-equilibrium ionization plasma model 
are shown for each cases. The error bars represent $1\sigma$ uncertainties.} 
\label{snr_spec}
\end{figure*}

We utilized the tool \emph{specextract} 
to extract the spectra and to compute the
response files. After background subtraction, there are 
net counts of $5118\pm73$~cts (outer rim), $1302\pm37$~cts (inner rim) and 
$1157\pm36$~cts (central region) available for analysis. According 
to the photon statistics, we group each of the extracted spectrum dynamically 
so as to achieve a comparable signal-to-noise ratio. 
The energy spectra of different 
components are displayed in Figure~\ref{snr_spec}.
Emission lines from various metals can be
clearly observed (e.g. Mg at $\sim1.4$~keV; Si at $\sim1.9$~keV), 
which support the SNR nature of \G. 

We examined these 
spectra with an absorbed non-equilibrium ionization model of 
a constant temperature and a single ionization timescale
(XSPEC model: NEI). All the spectral fittings 
were performed with XSPEC 12.6.0. The best-fit parameters 
are summarized in Table~\ref{spec_par}. 
All quoted errors are $1\sigma$ for 1 parameter of interest. 
For all three investigated 
regions, we do not find any conclusive evidence for the deviation of metal abundances 
from the solar values. 

The ionization timescale inferred from fitting the outer rim spectrum 
and the inner rim spectrum is 
$\tau_{\rm ion}=\left(7.3\pm1.5\right)\times10^{10}$~s~cm$^{-3}$ and 
$\tau_{\rm ion}=\left(4.3^{+2.1}_{-1.2}\right)\times10^{10}$~s~cm$^{-3}$ respectively, 
which indicate these parts of the remnant might not yet reach the collisional 
ionization equilibrium (CIE). The best-fit plasma temperature of outer rim and inner rim 
are found to be $kT=0.63^{+0.05}_{-0.02}$~keV and $kT=0.97^{+0.19}_{-0.16}$~keV respectively.
The higher temperature at the inner rim is consistent with the hardness distribution 
shown in the color image (i.e. Fig.~\ref{rgb}). The moderate temperature variation 
can possibly be caused by the complex density structure in the shocked region 
(e.g. see Chevalier 1982). 

For the spectrum of the central region,
the ionization timescale cannot be properly
constrained in NEI model fitting. We put a lower bound of $\tau_{\rm ion}$
at $\gtrsim10^{13}$~s~cm$^{-3}$. This suggests
the central region can possibly reach the condition for CIE already. 
The best-fit plasma temperature of this region is found to be $kT=0.57\pm0.03$~keV.

We further investigated if there is any non-thermal emission in various 
regions by adding a power-law model on the aforementioned best-fit plasma 
models. We found that the parameters of the additional power-law 
cannot be properly constrained. This prompts us to perform the 
spectral fit with the photon index fixed at the value of $\Gamma=2$. 
For all three spectra, no significant improvement of the goodness-of-fit 
have been found. We place $1\sigma$ upper limits of any non-thermal emission 
at the levels of $<1.9\times10^{-14}$~erg~cm$^{-2}$~s$^{-1}$, 
$<3.5\times10^{-14}$~erg~cm$^{-2}$~s$^{-1}$ and 
$<4.5\times10^{-14}$~erg~cm$^{-2}$~s$^{-1}$ for the outer rim, inner rim and the 
central region respectively. 

We have also examined the spectrum for the brightest source detected in the 
FOV (i.e. source \#7). The source spectrum has been extracted from a circular 
region with a radius of $5^{"}$ centered at the position reported by the source 
detection algorithm (cf. Tab~\ref{x_src}). For the background subtraction, we 
sampled from an annulus centered on the same position with an inner/outer radius
of $7^{"}$/$14^{"}$. There are $156\pm13$ net counts available for the spectral 
fitting. 

We have firstly examined the spectrum with various single component models, 
including the blackbody, the power-law and the hot diffuse gas model based 
on Mewe et al. (1985) (XSPEC model: MEKAL). 
We found all the tested single component model cannot 
provide any adequate description of the data. All result in a reduced \chisq\ 
(i.e. \chisq/d.o.f.) larger than 2 which indicate these models do not 
provide an adequate description of the data. We proceeded to perform the fitting 
with double blackbody which generally describes the spectrum of a particular 
manifestation of neutron stars, namely the central compact objects (CCOs) 
(Hui 2007; Mereghetti 2011). 

\begin{figure*}[t]
\centerline{\psfig{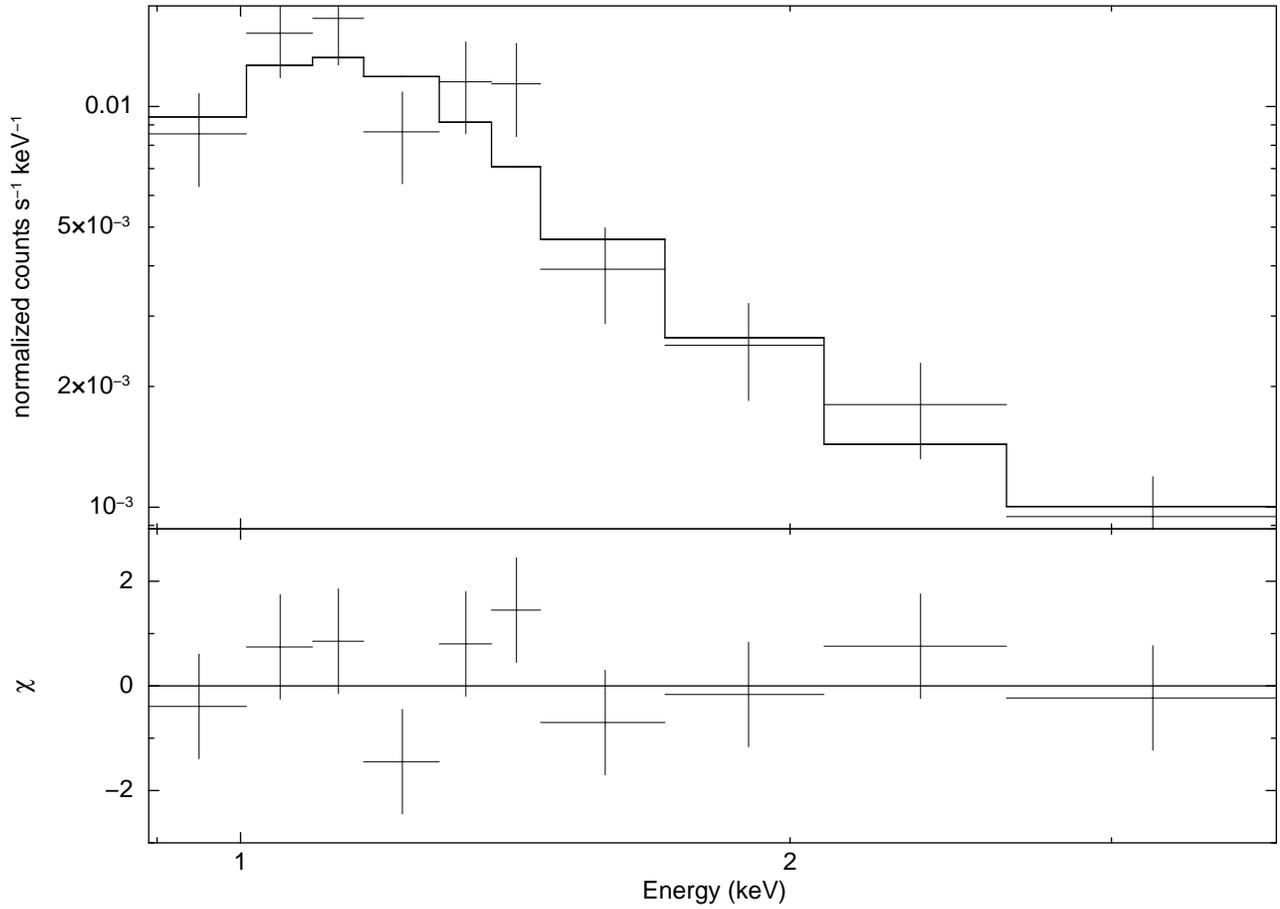}}
\caption[]{X-ray spectrum of the emission from the position of source \#7 as 
observed with ACIS-I with the best-fit double blackbody model (\emph{upper panel}) 
and contributions to the \chisq\ statistics (\emph{lower panel}). 
The error bars represent $1\sigma$ uncertainties.}
\label{cco_spec}
\end{figure*}

A double blackbody fit yields $N_{H}=(1.5\pm0.7)\times10^{22}$~cm$^{-2}$, 
$kT_{1}=0.09^{+0.04}_{-0.02}$~keV, $kT_{2}=0.44^{+0.21}_{-0.10}$~keV 
with an acceptable goodness-of-fit \chisq=7.3 for 5 d.o.f.. 
Their normalizations imply the emitting regions with the radii of
$R_{1}=27^{+30}_{-21}D_{\rm kpc}$~km and $R_{2}=35^{+34}_{-18}D_{\rm kpc}$~m, 
where $D_{\rm kpc}$ is the source distance in unit of kpc. 
The unabsorbed flux of source \#7 is found to be 
$f_{x}\simeq1.3\times10^{-11}$~erg~cm$^{-2}$~s$^{-1}$ in 0.5-8~keV. 
The comparison between this best-fit model and the observed data is 
displayed in Figure~\ref{cco_spec}. 

We have also examined the spectrum of source \#7 with a blackbody plus 
power-law model, which is a typical phenomenological description for the 
spectrum of a quiescent low-mass X-ray binary. This model yields 
$N_{H}=(1.5^{+1.2}_{-0.7})\times10^{22}$~cm$^{-2}$,
$kT=0.09^{+0.02}_{-0.03}$~keV, $R_{\rm bb}=24^{+23}_{-10}D_{\rm kpc}$~km, 
$\Gamma=2.3^{+0.7}_{-0.2}$ and the power-law normalization of 
$(6.0^{+4.7}_{-2.8})\times10^{-5}$ 
photons~keV$^{-1}$~cm$^{-2}$~s$^{-1}$ at 1 keV. The goodness-of-fit 
(\chisq=8.0 for 5 d.o.f.) is 
comparable with that resulted from the double blackbody fit. 
The unabsorbed flux in 0.5-8~keV inferred from this model is 
$f_{x}\simeq1.1\times10^{-11}$~erg~cm$^{-2}$~s$^{-1}$. 

For source\#5, the second brightest objects detected in the FOV without 
any optical counterpart identified, the 
source spectrum and the adopted background are extracted from the regions 
with the same sizes in the case of source \#7 but centered at the nominal 
position of source\#5 (Tab~\ref{x_src}). After background subtraction, there 
are $61\pm8$~cts available for the analysis. Among the tested single component 
model, we found that MEKAL provides a better description for the 
spectrum of this object (\chisq=1.4 for 3 d.o.f.) than a blackbody 
(\chisq=21.1 for 3 d.o.f.) or a power-law (\chisq=4.8 for 3 d.o.f.). 
The best-fit model yields $N_{H}=(5.7^{+2.9}_{-1.9})\times10^{21}$~cm$^{-2}$ 
and $kT=0.46^{+0.16}_{-0.26}$~keV. This infers an unabsorbed flux of 
$f_{x}\simeq2.0\times10^{-13}$~erg~cm$^{-2}$~s$^{-1}$ in 0.5-8~keV. 

\section{OPTICAL/INFRARED SPECTRAL ENERGY DISTRIBUTION OF THE CENTRAL COMPACT OBJECT}

We have also examined the optical properties for the counterpart of source \#7. 
In USNO-B1.0 catalog, only a single source with the magnitudes of $B=19.72$, $R=19.64$ 
and $I=16.15$ has been identified within its $10\sigma\times10\sigma$ X-ray error box. 
We have also searched for the infrared counterpart in the Two Micron All Sky Survey 
(2MASS) catalog (Skrutskie et al. 2006). Similarly, only one source with $J=14.01$, $H=13.31$, 
and $K_{s}=13.10$ can be found with its error box. 
To complement these catalog values, we have also carried out observations in $V$ and $U$ band so 
as to have a more complete frequency coverage. 

For $U$ band, 
by applying aperture photometry to the stacked UVOT image with an exposure of 6.49~ks, we have placed an limiting 
magnitude of 21.29. The UVOT data also allow us to cross-check the magnitude in $B$ band, from which
we have obtained $B$=19.77 which confirms the value reported in USNO-B1.0 catalog. 
For $V-$band, we have observed source \#7 with the Cerro Tololo Interamerican
Observatory (CTIO). Data have been obtained using the ANDICAM dual-channel imaging photometer
at the SMARTS/CTIO 1.3~m telescope in $V$ filter. The zero-point corrected magnitude is found to be 
$V=19.16$. 

\begin{figure*}[t]
\centerline{\psfig{figure=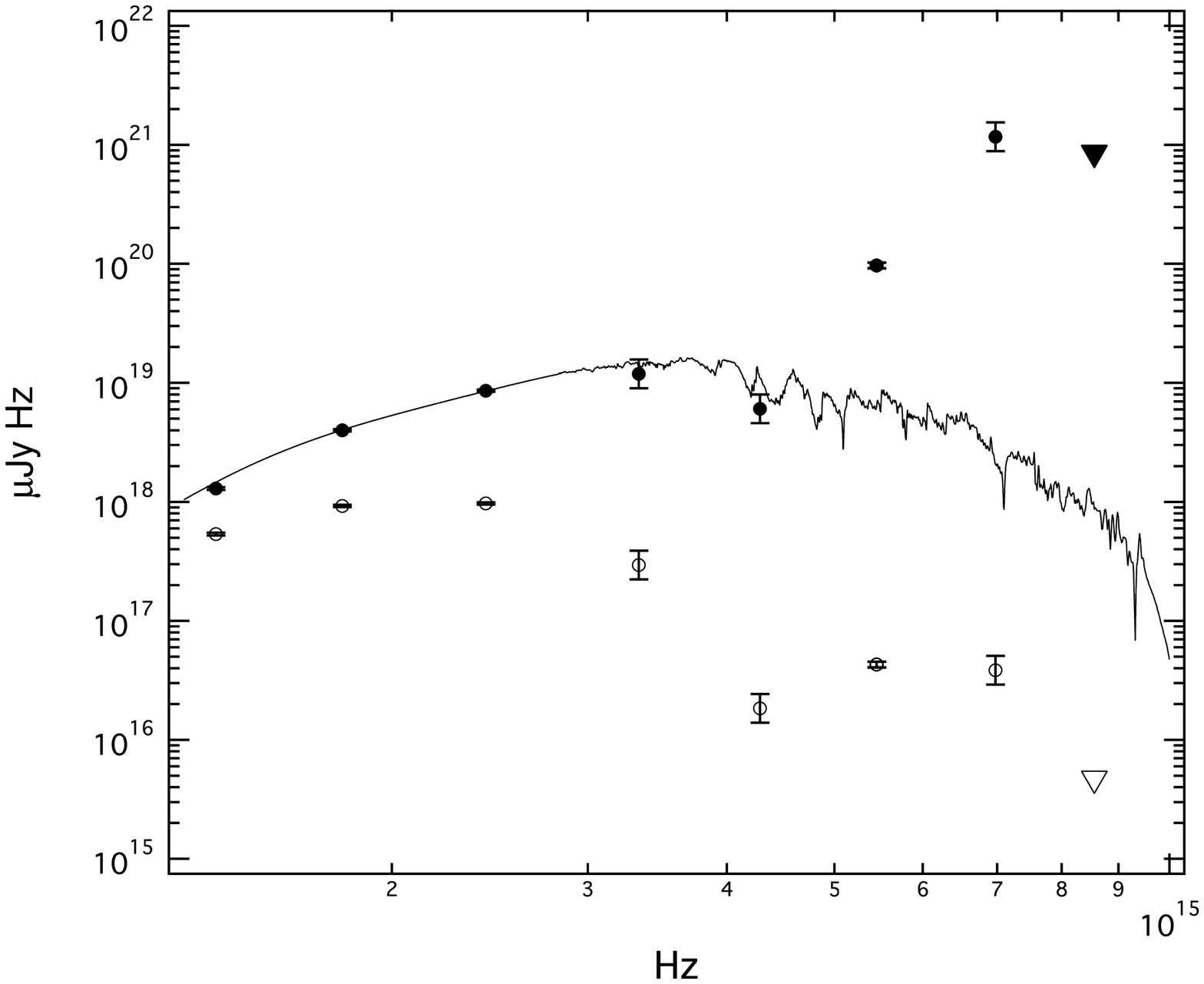,width=18cm,clip=,angle=0}}
\caption[]{Optical/infrared spectral energy distribution of source \#7. Both observed (open symbols) 
and de-reddened (solid symbols) data points are shown in this plot. The open and solid triangles represent the 
observed and de-reddened $3\sigma$ upper limit inferred from SWIFT UVOT observation. 
The spectral model of a M3V star obtained from the stellar spectral flux library (Pickles 1998) is overplotted. 
The error bars represent the photometric uncertainties corresponding to each data point.}
\label{oir_sed}
\end{figure*}

These identifications enable us to construct an optical/infrared 
spectral energy distribution (SED) which is shown in Figure~\ref{oir_sed}.  
We adopted the column density inferred the X-ray spectral fit of source \#7 
(i.e. $1.5\times10^{22}$~cm$^{-2}$) to perform the extinction-correction 
(cf. Predehl \& Schmitt 1995; Cardelli, Clayton, \& Mathis 1989). The de-reddened 
SED has also been plotted in Figure~\ref{oir_sed}. For the range from $K_{s}$ band 
to $R$ band, the SED can be modeled with the spectrum of a M3V star. On the other hand,
an apparent excess is found at and beyond $V$ band. 

To further constrain its optical properties, 
we have also searched for its H$\alpha$ counterpart from the SuperCOSMOS H-alpha 
Survey (SHS) catalog (Parker et al. 2005). Within the X-ray error box of source \#7, 
only one source with the magnitude 17.35 is found. In comparison with $R$ band, $m_{\rm H\alpha}-m_{R}=-2.3$, 
a clear enhancement in H$\alpha$ is noted which suggests the presence of a strong emission line.

\begin{table*}[b]
\caption{X-ray spectral parameters inferred from different regions in \G.}
\label{spec_par}
\begin{center}
 \begin{tabular}{l |c | c | c}
  \hline\hline\\[-2ex]
  & Outer rim &  Inner rim & Central region \\\\[-2ex]
\hline\\[-2ex]
 $N_{H}$ ($10^{21}$ cm$^{-2}$)  & $10.3^{+0.3}_{-0.4}$ & $9.7^{+0.8}_{-0.7}$ & $9.4^{+0.3}_{-0.4}$ \\[1ex]  
 $kT$ (keV) & $0.63^{+0.05}_{-0.02}$ & $0.97^{+0.19}_{-0.16}$ & $0.57\pm0.03$ \\[1ex] 
 $\tau_{\rm ion}$ (s~cm$^{-3}$)  & $(7.3\pm1.5)\times10^{10}$ & $(4.3^{+2.1}_{-1.2})\times10^{10}$ 
& $>10^{13}$ \\[1ex]
 Norm  ($10^{-3}$)$^{a}$  & $4.5^{+0.6}_{-0.7}$ & $0.5^{+0.2}_{-0.1}$ & $0.9\pm0.1$ \\[1ex]  
\hline\\[-2ex]
 $f_{x}$ ($10^{-12}$~erg~cm$^{-2}$~s$^{-1}$)$^{b}$ & $24.1^{+3.1}_{-3.4}$ & $3.8^{+1.1}_{-0.9}$ & $2.3^{+0.4}_{-0.3}$ \\[1ex]  
\hline\\[-2ex]
\chisq\                   & 157.08  & 141.49  & 131.94 \\[1ex]
D.O.F.                    & 138  & 129  & 122 \\[1ex]
 \hline
 \end{tabular}
 \end{center}
$^{a}$ {\footnotesize The model normalization is expressed as
$(10^{-14}/4\pi D^{2})\int n_{e}n_{H}dV$ where $D$ is the source distance
in cm and $n_{e}$ and $n_{\rm H}$ are the post-shock electron and hydrogen densities in cm$^{-3}$.}\\
$^{b}$ {\footnotesize Unabsorbed flux over the energy range of $0.5-8$~keV.}\\
 \end{table*}

\section{DISCUSSION}

We have performed a detailed spectro-imaging X-ray study of the SNR candidate \G\ with
Chandra. An incomplete shell-like X-ray structure, which is well-correlated with radio shell, 
has been revealed. Its X-ray spectrum has shown the presence of a hot plasma accompanied with 
metallic emission lines. 
All these observational evidences clearly suggest \G\ is indeed a SNR.

Utilizing the X-ray results, we proceed to discuss the properties of \G. 
The emission measure inferred from the spectral fits of the outer rim 
emission allows us to estimate the hydrogen density $n_{\rm H}$ and the electron density $n_{\rm e}$
in the shocked regions. Assuming the shocked densities of hydrogen and
electrons are uniform in the extraction region, the normalization
of the NEI model can be approximated by $10^{-14}n_{\rm H}n_{e}V/4\pi D^{2}$, 
where $D$ is the distance
to \G\ in cm and $V$ is the volume of interest in units of cm$^{3}$.
Assuming a geometry of oblated spheroid, the volume of interest for the outer rim 
is  $5\times10^{54}D^{3}_{\rm kpc}$ cm$^{3}$ where $D_{\rm kpc}$ is the remnant distance 
in units of kpc. Assuming a fully ionized plasma with $\sim10\%$ He ($n_{\rm e}\sim1.2n_{\rm H}$), 
the $1\sigma$ confidence interval of the outer rim normalization implies the shocked hydrogen 
and electron densities in the ranges of $n_{\rm H}\simeq\left(2.8-3.2\right)D^{-0.5}_{\rm kpc}$~cm$^{-3}$  
and $n_{\rm e}\simeq\left(3.3-3.9\right)D^{-0.5}_{\rm kpc}$~cm$^{-3}$ respectively. 

Assuming \G\ is in a Sedov phase, the shock temperature can be estimated by
$T_{s}\simeq 8.1\times 10^{6}E_{51}^{2/5}n_{\rm ISM_{-1}}^{-2/5}t_{4}^{-6/5}$~K,
where $t_{4}$, $E_{51}$ and $n_{\rm ISM_{-1}}$ are the time after 
the explosion in units of $10^{4}$~years,
the released kinetic energy in units of $10^{51}$~ergs and the ISM 
density of 0.1~cm$^{-3}$ respectively. Assuming it is a strong shock, $n_{\rm ISM}$ is estimated 
as $0.25n_{\rm H}$. Taking the $1\sigma$ uncertainties of the temperature and the normalization 
for the outer rim into account (cf. Tab.~\ref{spec_par}), the age of \G\ is constrained in the 
brackets of $t\simeq\left(2.4-2.7\right)D_{\rm kpc}^{1/6}\times10^{3}$~yrs and  
$t\simeq\left(5.1-5.9\right)D_{\rm kpc}^{1/6}\times10^{3}$~yrs for $E_{51}=0.1$ and $E_{51}=1$ respectively.
Since the distance plays a crucial role in determining the physical properties of \G,
a follow-up HI observation of \G\ is strongly recommended.

Our observation has also revealed a number of sources in the field of \G\ (cf. Fig.~\ref{wavelet}).
Source \#7 is the most interesting source in this population. First, among all the detected sources, 
it is the closest object from the geometric center of the remnant with an offset of $\sim1.4^{'}$. 
Also, the column absorption of this object inferred from the spectral fit is not far from 
the value inferred from the remnant spectrum, which suggests a possible tie between 
source \#7 and \G. Its spectrum can be described by a composite blackbody or a blackbody plus power-law 
model, though the parameters  
are rather unconstrained due to the small photon statistic resulted from this short exposure. We would 
like to point out that these properties are similar to those of CCOs, which is 
one of the most poorly known classes among all known manifestations of neutron stars (see Mereghetti 2011 
for a recent review). CCOs are typically characterized by their proximity to the expansion center of 
the associated SNRs as well as their thermal X-ray spectra (Mereghetti 2011; Hui 2007). 

On the other hand, all the known CCOs are also characterized by their large X-ray-to-optical flux 
ratios ($f_{x}/f_{\rm opt}>10^{3}$) which is typical for isolated neutron stars (cf. De Luca 2008). 
In searching for the counterpart of source \#7, we have identified a single optical/infrared source 
within the error box of source \#7, which clearly rules out the possibility of 
an isolated neutron star. 

However, we cannot exclude the possibility that this is a neutron star resides in a binary 
system that survived in the SN explosion. The binary scenario is also suggested by the putative modulation 
of $P\sim1.4$~hrs, which can possibly be the orbital period of this system. A similar scenario has also been 
suggested to explain the X-ray temporal behavior of the peculiar 
CCO associated with the SNR RCW~103 which has a periodic modulation at 6.67~hrs that can also be 
interpreted as the orbital period (Pizzolato et al. 2008; De Luca et al. 2006). Nevertheless, this 
relatively short exposure in our observation do not provide a firm conclusion on the periodicity of source \#7. 
A longer follow-up X-ray observation can enable us to better constrain its temporal behavior. 

Similar to the CCO in RCW~103, flux variability has also been detected from source \#7 in \G\ (see 
Fig.~\ref{src7_var}). The nature of this variability remains unknown. Speculations have been ranged from 
the instability of the accretion disc around the compact object to the scenario that the CCO is a 
magnetar (De Luca et al. 2006). The X-ray source in RCW~103 is so far the only CCO that has demonstrated 
long-term flux variability. It is possible that source \#7 in \G\ can be the second example. Multi-epoch 
X-ray monitoring of this source can provide a deeper insight on its temporal behavior. 

The properties of the infrared/optical counterpart of source \#7 are also worth discussing. We found that its
SED can be well-described by the model spectrum of a late-type star in the range from $K_{s}$ band to $R$ band 
(see Fig.~\ref{oir_sed}). In the context of a binary scenario, this suggests the evidence for the companion 
star of source \#7. Nevertheless, with these sparse data points obtained from the catalogs, its nature cannot 
be constrained unambiguously. Dedicated spectroscopic observations are required to provide us a deeper insight  
of it. 

The other interesting optical properties are the enhancements observed in $V$, $B$ bands as well 
as H$\alpha$. These characteristics suggest the presence of an accretion disk around a compact object 
and a low mass companion (e.g. Cool et al. 1998). 
However, these excesses are larger than those observed from a typical quiescent 
low-mass x-ray binary (e.g. Haggard et al. 2004). As the orbital period of source \#7 is suggested to be 
$\sim1.4$~hrs, it is expected to be a very compact binary system. Such tight orbit can possibly lead to the 
aforementioned large enhancements. Confirmation of this putative orbital period can provide a natural
explanation of the optical properties. 

We would like to highlight that there are only 11 CCOs have been uncovered so far 
(cf. Tab.~1 in Mereghetti 2011). With this limited sample size, we cannot even determine 
if these objects consist of a homogeneous class. For a better understanding of their nature, 
the sample of CCOs has to be enlarged. In this aspect, our long-term 
X-ray identification campaign of SNR candidates 
can also enable the search for the candidates of these compact objects. 

\section{SUMMARY}
There is a group of unidentified extended RASS objects have been suggested as promising SNR candidates. 
We have initiated a long-term identification campaign by observing the brightest candidate, \G, with 
Chandra. With a short exposure, we have confirmed the nature of \G\ as a SNR through a detailed X-ray 
spectro-imaging analysis. Apart from the remnant emission, a bright X-ray point source locates close to
the geometrical center of \G\ has also been detected as a new CCO. This compact object has shown a 
putative periodicity of $\sim1.4$~hrs and excesses in $V$, $B$ bands and H$\alpha$, which suggest 
it as a promising candidate of a compact binary survived in a SN. In conclusion,
this proposed alternative
window of SNRs and compact objects survey has been demonstrated to be fruitful by our pilot target. 

\acknowledgments{
The authors would like to thank the anonymous referee for the useful comments. 
CYH is supported by the National Research Foundation of Korea through grant 2011-0023383. 
LT would like to thank the German Deutsche Forschungsgemeinschaft, DFG for financial support in project 
SFB TR 7 Gravitational Wave Astronomy.
AKHK is supported partly by the National 
Science Council of the Republic of China (Taiwan)
through grant NSC99-2112-M-007-004-MY3 and a Kenda
Foundation Golden Jade Fellowship. 
}


\end{document}